# Event Occurrence Model for Power Distribution System Based on Data Eigenvalue Behavior Analytics

Nan Zhou[a], Lingen Luo[a,], Gehao Sheng[a], Xiuchen Jiang[a]

[a] Department of Electrical Engineering, Shanghai Jiao Tong University, No. 800 Dongchuan Road, 200240,

Shanghai, People's Republic of China

## I. Power Distribution Network Modelling

### A. Power Distribution Network PF Model

We limit our discussion to a single-phase power distribution network which is modeled with an undirected graph denoted by $\mathcal{G} = (\Upsilon, \mathcal{E})$ where [1], [2]:

- $\Upsilon$ is the set of buses or nodes, with size $N+1$;
- $\mathcal{E}$ is the set of electrical lines with size $m$ and $m \geq N$;

During operation state, the distribution network is often assumed to be lossless, which means the power generation minus demand is zero. In that case, one node in the network can be considered as the reference node and the voltage and phase at the reference node is considered at a fixed reference value without a loss of generality. In this work, the substation node is selected as the reference node. The substation is modeled as an ideal voltage generator, be labeled as node 0 and let $v_0$, $i_0$ be its nodal voltage and injected current respectively. The state of the power distribution network can be described by the nodal voltage vector $v \in \mathbb{C}^{N+1}$ and the injected current vector $i \in \mathbb{C}^{N+1}$. Let $z_{ab} = r_{ab} + \hat{i} \cdot x_{ab}$ ($\hat{i}^2 = -1$) be the complex line impedances of line ($ab \in \mathcal{E}, a \neq b$). Derived from the Kirchhoff laws, the single-phase PF model of complex powers and potentials (voltages and phases) at node $a$ is given as:

$$\forall a \in \Upsilon : P_a = p_a + \hat{i} \cdot q_a = \sum_{b: \forall (a,b) \in \mathcal{E}} \frac{V_a(V_a^* - V_b^*)}{z_{ab}^*} = \sum_{b: \forall (a,b) \in \mathcal{E}} \frac{v_a^2 - v_a v_b \cdot \exp(\hat{i} \cdot \theta_a - \hat{i} \cdot \theta_b)}{z_{ab}^*} \quad (1)$$

where $p_a$, $q_a$, $v_a$ and $\theta_a$ are respectively the active and reactive power injection, voltage magnitude and phase at node $a$. $V_a = v_a \cdot \exp(\hat{i} \cdot \theta_a)$. $V_b$, $v_b$ and $\theta_b$ are those of node b. $\exp(\cdot)$ stands for the exponential function with natural constant $e$ as base. $(\cdot)^*$ stands for complex conjugate.

We assume that the phase difference between neighboring nodes ($\theta_a - \theta_b$) and deviations of the voltage magnitude ($v_a - 1$) from the reference node of 1 p.u. are both small [3], [4]. This means $\sin(\theta_a - \theta_b)$ is approximately equal to ($\theta_a - \theta_b$), $\cos(\theta_a - \theta_b)$ is approximately equal to 1 and $v_a$ is approximately 1. Apply Euler's formula to $\exp(\hat{i} \cdot \theta_a - \hat{i} \cdot \theta_b)$, we have:

$$\exp(\hat{i} \cdot \theta_a - \hat{i} \cdot \theta_b) = \cos(\theta_a - \theta_b) + \hat{i} \cdot \sin(\theta_a - \theta_b) \quad (2)$$

Substitute (S2) into (S1) and we have:

$$P_a = p_a + \hat{i} \cdot q_a = \sum_{b: \forall (a,b) \in \mathcal{E}} \frac{v_a^2 - v_a v_b \cdot \left( \cos(\theta_a - \theta_b) + \hat{i} \cdot \sin(\theta_a - \theta_b) \right)}{z_{ab}^*} \quad (3)$$

Multiply the conjugate complex of the denominator, we have:

$$\begin{aligned} P_a = p_a + \hat{i} \cdot q_a &= \sum_{b: \forall (a,b) \in \mathcal{E}} \frac{\left[ v_a^2 - v_a v_b \cdot \left( \cos(\theta_a - \theta_b) + \hat{i} \cdot \sin(\theta_a - \theta_b) \right) \right] \cdot z_{ab}}{z_{ab}^* \cdot z_{ab}} \\ &= \sum_{b: \forall (a,b) \in \mathcal{E}} \left[ \frac{r_{ab}(v_a^2 - v_a v_b \cos(\theta_a - \theta_b)) + x_{ab} v_a v_b \sin(\theta_a - \theta_b)}{r_{ab}^2 + x_{ab}^2} + \hat{i} \cdot \frac{x_{ab}(v_a^2 - v_a v_b \cos(\theta_a - \theta_b)) - r_{ab} v_a v_b \sin(\theta_a - \theta_b)}{r_{ab}^2 + x_{ab}^2} \right] \end{aligned} \quad (4)$$

Substitute the approximations into the relation above, we eventually obtain the PF model as:

$$\begin{aligned} p_a &= \sum_{b: \forall (a,b) \in \mathcal{E}} \left( g_{ab}(v_a - v_b) + \beta_{ab}(\theta_a - \theta_b) \right) \\ q_a &= \sum_{b: \forall (a,b) \in \mathcal{E}} \left( \beta_{ab}(v_a - v_b) - g_{ab}(\theta_a - \theta_b) \right) \end{aligned} \quad (5)$$

where $g_{ab} \doteq \frac{r_{ab}}{r_{ab}^2 + x_{ab}^2}$ and $\beta_{ab} \doteq \frac{x_{ab}}{r_{ab}^2 + x_{ab}^2}$.

The relation (5) can be represented in the matrix form as:

$$P = M^T gM \cdot v + M^T \beta M \cdot \theta$$
$$Q = M^T \beta M \cdot v - M^T gM \cdot \theta \quad (6)$$

It can also be represented in the complex matrix form as (which is equation (3) in the original text):

$$P + \hat{i}Q = M^T (g + \hat{i}\beta) M (v - \hat{i}\theta) \quad (7)$$

where $M$ is the edge to node incidence matrix: every row $m_{ab}$ in $M$ is equal to $(e_a^T - e_b^T)$ which represents the edge $(a, b)$. $e_a$ is the standard basis vector with 1 at the $a^{th}$ position and zero elsewhere. $g$ and $\beta$ are both diagonal matrixes representing their respective line conductance and susceptance. An illustration of the network incidence matrix, diagonal matrix and conductance matrix is shown in Fig. 1.

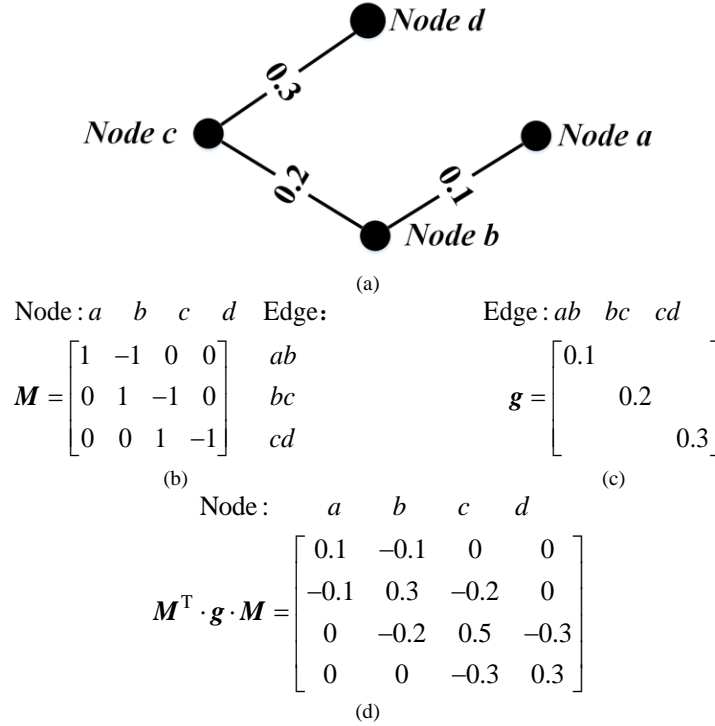

Fig. 1. Illustration of the network incidence matrix establishment on the example of a network with four nodes (a) a network with four nodes ($a, b, c, d$). The line conductance is as written in the figure. (b) the node incidence matrix (c) diagonal matrix representing line conductance (d) the network conductance matrix.

Note before eliminating the reference node, the matrixes in (6) and (7) are ($N$+1)-dimensional matrixes with rank $N$. This means these matrixes are linearly dependent and are noninvertible. Therefore, by eliminating the substation node, which means eliminating the corresponding vectors from the matrixes, the resulting reduced matrixes are full-rank and invertible. Take the matrix $M$ in Fig. 1 as example, eliminating the vector of Node a makes the previously noninvertible matrix invertible. Eliminate the reference node from all the matrixes in (6) and (7). For simplicity we use the same notation for the original and reduced dimension matrices.

By inverting the relation in (6), we can express the voltage magnitude and phase in the form of active and reactive power as:

$$\theta = M^{-1} x (M^{-1})^T p - M^{-1} r (M^{-1})^T q$$
$$v = M^{-1} r (M^{-1})^T p + M^{-1} x (M^{-1})^T q \quad (8)$$

with $r$ and $x$ both diagonal matrixes with their element as the line resistance and reactance.

The relation in (8) is the equation (4) in the original text.

*B. Covariance Matrix of Nodal Voltage Matrix*

Based on the derivation above, our next aim is to obtain the voltage eigenvalue spectrum, which should be calculated from its covariance matrix. Denote the mean value of complex variable $x$ as $\mu_x$. The auto-covariance matrix of $x$ writes as:

$$\Omega_x = E[(x - \mu_x)(x - \mu_x)^*] \quad (9)$$

where $(\cdot)^*$ denotes conjugate transpose. Moreover, denote the mean value of complex variable $y$ as $\mu_y$. The cross-covariance matrix of $x$ and $y$ writes as:

$$\Omega_{xy} = E[(x - \mu_x)(y - \mu_y)^*] \quad (10)$$

Thus, the covariance matrix of phase difference $\theta$ and voltage amplitude $v$ (denoted as $\Omega_v$) can be written as:

$$\Omega_\theta = E[(\theta - \mu_\theta)(\theta - \mu_\theta)^*] = H_{1/x}^{-1}\Omega_p H_{1/x}^{-1} + H_{1/r}^{-1}\Omega_q H_{1/r}^{-1} - H_{1/x}^{-1}\Omega_{pq} H_{1/r}^{-1} - H_{1/r}^{-1}\Omega_{qp} H_{1/x}^{-1}$$

$$\Omega_v = E[(v - \mu_v)(v - \mu_v)^*] = H_{1/r}^{-1}\Omega_p H_{1/r}^{-1} + H_{1/x}^{-1}\Omega_q H_{1/x}^{-1} + H_{1/r}^{-1}\Omega_{pq} H_{1/x}^{-1} + H_{1/x}^{-1}\Omega_{qp} H_{1/r}^{-1}$$

(11)

where $H_{1/r} = M^T r^{-1} M$ and $H_{1/x} = M^T x^{-1} M$.

## II. EVENT OCCURRENCE DETECTION AND CLASSIFICATION METHOD BASED ON EIGENVALUE DISTRIBUTION ANALYSIS

### A. Event Occurrence Model

Suppose an event occurred at node $k$ in the distribution system in Fig. 2. The nodal voltage, injected current and the line impedance ($z_i$, $i$=1, 2, ..., m) are as labelled in the figure.

According to the compensation theory, once an element change occurred in the circuit, it can be replaced with a current source which injects the same amount of current as the current changes through the element. Therefore, we replace the element change at node $k$ in Fig. 2 with a current source $\Delta i_k = I^{post} - I^{pre}$, where $I^{pre}$ and $I^{post}$ denotes respectively the current drawn by the element before and after the event.

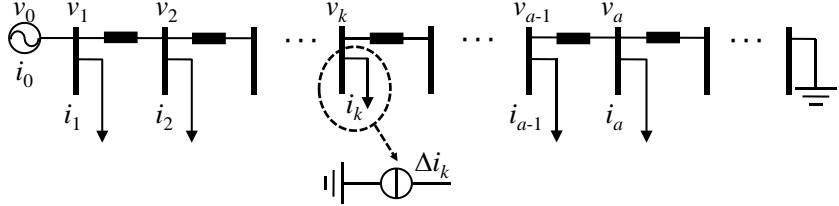

**Fig. 2.** The illustration of a distribution system with the reference node denoted as node 0 and the equivalent circuit based on compensation theorem.

### B. Voltage Covariance Matrix Reacting to Event Occurrence

This subsection gives the detailed derivation how the event occurrence influences the eigenvalue distribution of $\Omega_v$. The influence of the event occurrence to the distribution system nodal voltage matrix can be summarized as below:

1) Once an event occurs in the distribution network, it can be replaced with a current source $\Delta i_k$ at node $k$. The current deviation $\Delta I$ writes as $\Delta I_k = \Delta I + \Delta i_k$.

2) Through (9), this current deviation $\Delta I_k$ caused a perturbation $\Delta\Omega_P$ into the power covariance matrix $\Omega_P$.

3) Through (7), this perturbation $\Delta\Omega_P$ caused by the injected current source is eventually transferred into the nodal voltage covariance matrix $\Omega_v(a, a)$. This makes the eigenvalues of $\Omega_v$ exceeds the limit of SR law and M-P law, as shown in Fig. 3.

To demonstrate the conclusion above, we divided our derivation into two parts: (i) we first verified that the power covariance is related to the current covariance; (ii) we then demonstrated that the voltage covariance is related to the power covariance.

Part (i): Denote the network voltage matrix as $V$, current matrix as $I$ and admittance matrix as $Y$. We write the network relation before event occurrence as:

$$V \cdot Y = I \quad (12)$$

There is a linear relation between voltage deviation $\Delta V$ and power deviation $\Delta P$ which is expressed as $\Delta P = J_V \cdot \Delta V$ where $J_V$ is the Jacobian matrix at the voltage matrix mean value. Substitute this relation into the network relation, we have $\Delta P = J_V Y^{-1} \Delta I$.

The covariance of power matrix $\Omega_P$ can be obtained as:

$$\Omega_P = E[(P - \mu_P)(P - \mu_P)^*] = E[\Delta P \Delta P^*] = E[Y^{-1} J_V \Delta I \cdot \Delta I^* J_V^* (Y^{-1})^*]$$
$$= Y^{-1} J_V E[\Delta I \cdot \Delta I^*] J_V^* (Y^{-1})^* = Y^{-1} J_V \Omega_I J_V^* (Y^{-1})^*$$

(13)

Part (ii): We first introduce the following assumption:

***Assumption 1:*** The active and reactive power demands of the individual nodes are uncorrelated, while active and reactive powers at the same node are positively correlated.

This assumption is a simplification reflecting the diversity of individual demands in the power distribution network. Thus, for any node $\forall a, b \in \Upsilon$, there is no correlation between different nodes' power, which means the cross-covariance matrix is zero. In mathematics, this means:

$$\Omega_p(a,b) = \Omega_q(a,b) = \Omega_{pq}(a,b) = 0 \quad (14)$$

Note there is a difference between the covariance matrix $\Omega_p$ and $\Omega_P(a, b)$. The former is calculated with (6) which is the active power matrix of all nodes in system, while the latter is calculated with only the active power from node $a$ and node $b$. This feature also applies to the covariance matrix of $\Omega_q(a, b)$ and $\Omega_{pq}(a, b)$.

Thus, for any node $\forall a \neq b$, the covariance of nodal voltage at node $a$ can also be represented with:

$$\Omega_v(a,a) = \Omega_v(a,a) - 0 = \Omega_v(a,a) - \Omega_v(a,b) \quad (15)$$

where both the auto-covariance matrix $\Omega_v(a,a)$ and cross-covariance matrix $\Omega_v(a,b)$ are in the form of (11) with four nonnegative terms. Let the first term $H_{1/r}^{-1}(a,a)\Omega_p(a,a)H_{1/r}^{-1}(a,a)$ denoted as $\Omega_v^1(a,a)$. The second, third and fourth term in (11) are denoted as $\Omega_v^2(a,a)$, $\Omega_v^3(a,a)$ and $\Omega_v^4(a,a)$ respectively. Following the similar process, we can write (15) as:

$$\Omega_v(a,a)-\Omega_v(a,b)=\left(\Omega_v^1(a,a)-\Omega_v^1(a,b)\right)+\left(\Omega_v^2(a,a)-\Omega_v^2(a,b)\right)+\left(\Omega_v^3(a,a)-\Omega_v^3(a,b)\right)+\left(\Omega_v^4(a,a)-\Omega_v^4(a,b)\right) \tag{16}$$

Equation (16) needs to be further simplified. To do so, we first give the following lemma:

**Lemma 1**: If node $a$ and its parent node $b$ both belong to the same tree $T_k$, we have:

$$\begin{aligned} H_{1/r}^{-1}(a,c) - H_{1/r}^{-1}(b,c) &= r_{ab} & ,c \in D_a^{T_k} \\ H_{1/x}^{-1}(a,c) - H_{1/x}^{-1}(b,c) &= x_{ab} & ,c \in D_a^{T_k} \end{aligned} \tag{17}$$

where $D_a^{T_k}$ is the set of nodes from $a$ to the reference node in tree $T_k$.

**Proof:** From the geometric relationship between the two nodes in the graph, we have:

$$D_a^{T_k} = D_b^{T_k} \cap \{(ab)\} \tag{18}$$

Recall the definition of the matrix $H_{1/r}$, $H_{1/x}$ and $M$ in (7) and (11). Note the difference between $H_{1/r}$ and $H_{1/r}(a,b)$ where the former is the matrix established with all nodes in the tree where the latter is established with only two nodes $a$ and $b$. Following the illustration in Fig. 1, equation (17) can be proved.

**Lemma 2**: If node $a$ and node $b$ are two nodes in the graph:

$$\begin{aligned} H_{1/r}^{-1}(a,b) &= 0 & ,\text{if } a,b \text{ are on different trees} \\ H_{1/r}^{-1}(a,b) &= \sum_{f \in T_k} M^{-1}(a,f) r(f,f) M^{-1}(b,f) \\ &= \sum_{(cd) \in D_a^{T_k} \cap D_b^{T_k}} r_{cd} & ,\text{if } a,b \in T_k \end{aligned} \tag{19}$$

Lemma 2 shows that $H_{1/r}^{-1}(a,b)$ represents the sum of the line resistances of the common lines of $D_a^{T_k}$ and $D_b^{T_k}$. If there is no common line, the result is 0. The same results can also be obtained with $H_{1/x}^{-1}(a,b)$.

Back to the matrix subtraction in (16) and apply Lemma 1. We can obtain:

$$\Omega_v^1(a,a)-\Omega_v^1(a,b)=H_{1/r}^{-1}(a,a)\Omega_p(a,a)H_{1/r}^{-1}(a,a) - H_{1/r}^{-1}(a,b)\Omega_p(a,b)H_{1/r}^{-1}(a,b) \tag{20}$$

Denote $d$ as the node in tree $T_k$ while $c$ is the node in $D_a^{T_k}$. Substitute $a$ with $d$ and then replace $b$ with $c$ in the relation above, we have:

$$\begin{aligned} \Omega_v^1(a,a)-\Omega_v^1(a,b) &= \sum_d \sum_{c \in D_a^{T_k}} H_{1/r}^{-1}(a,d)\Omega_p(d,c)\left(H_{1/r}^{-1}(a,c)-H_{1/r}^{-1}(d,c)\right) \\ &= \sum_d \sum_{c \in D_a^{T_k}} H_{1/r}^{-1}(a,d)\Omega_p(d,c) r_{ac} \\ &= \sum_{c \in D_a^{T_k}} H_{1/r}^{-1}(a,c)\Omega_p(c,c) r_{ac} \end{aligned} \tag{21}$$

Apply Lemma 2 to (21) and we finally have:

$$\Omega_v^1(a,a)-\Omega_v^1(a,b) = \sum_{\forall c \in D_a^{T_k}} r_{ac}^2 \cdot \Omega_p(c,c) \tag{22}$$

Following a similar procedure, we can obtain the expression of the other three terms:

$$\begin{aligned} \Omega_v^2(a,a)-\Omega_v^2(a,b) &= \sum_{\forall c \in D_a^{T_k}} x_{ac}^2 \cdot \Omega_p(c,c) \\ \Omega_v^3(a,a)-\Omega_v^3(a,b) &= \Omega_v^4(a,a)-\Omega_v^4(a,b) = \sum_{\forall c \in D_a^{T_k}} r_{ac} \cdot x_{ac} \cdot \Omega_p(c,c) \end{aligned} \tag{23}$$

Thus, the final expression of $\Omega_v(a,a)$ can be written as:

$$\Omega_v(a,a) = \sum_{\forall c \in D_a^{T_k}} [r_{ac}^2 \Omega_p(c,c) + x_{ac}^2 \Omega_q(c,c) + 2r_{ac} x_{ac} \Omega_{pq}(c,c)] \tag{24}$$

In (9), we have related the voltage covariance matrix $\Omega_v$ to the injected power covariance $\Omega_P$.

*C. Voltage Covariance Matrix Reacting to Event Occurrence*

Denote the network voltage matrix as $V = [v_1, v_2, \ldots, v_N]$ where $v_j = [v_{1j}, v_{2j}, \ldots, v_{Nj}]^T$ is a $N$-dimensional vector established with the $j^{th}$ column of $V$. The current matrix is in the similar form of $I = [i_1, i_2, \ldots, i_N]$ where $i_j = [i_{1j}, i_{2j}, \ldots, i_{Nj}]^T$. Denote the $N \times N$ dimensional impedance matrix as $Z$. We can write the network relation as:

$$V = Z \cdot I \quad (25)$$

Then convert the network voltage matrix $V$ and into its standard form $\tilde{V}$ with the method below:

$$\tilde{v}_j = v_j / [\sqrt{N} \cdot \sigma(v_j)], \quad j = 1, 2, \ldots, N \quad (26)$$

where $\sigma(\cdot)$ is the standard deviation and $\tilde{V} = [\tilde{v}_1, \tilde{v}_2, \ldots, \tilde{v}_N]$. Perform the similar conversion to the current matrix $I$ and eventually we have:

$$\tilde{V} = Z \cdot \tilde{I} + I_N \sigma_m^2 \quad (27)$$

where $I_N$ is the $N$ dimensional identity matrix and $\sigma_m^2$ denotes the Gaussian noise of the introduced by the measurement device.

The covariance matrix is a square matrix giving the covariance between each pair of elements of a given matrix. The covariance matrix of system nodal voltage generalizes the notion of variance to multiple dimensions, giving us the possibility to fully characterize different type of events in the system. Substitute the nodal voltage into (9), the voltage covariance can be obtained as:

$$\Omega_{\tilde{V}} = E[(\tilde{V} - \mu_{\tilde{V}})(\tilde{V} - \mu_{\tilde{V}})^*] = E[Z \cdot (\tilde{I} - \mu_{\tilde{I}})(\tilde{I} - \mu_{\tilde{I}})^* \cdot Z^*] + I_N \sigma_m^2 = Z \cdot Z^* + I_N \sigma_m^2 \quad (28)$$

Suppose an event occurred in the system. Because the event is replaced with an equivalent current source as described in the former section, it can be viewed as a sudden change in the $k^{\text{th}}$ entry of $\tilde{I}$. Let $\tilde{V}'$ denotes the system nodal voltage after event occurrence, the equation (27) can be written as:

$$\tilde{V}' = Z \cdot (I_N + \alpha_k e_k e_k^*) \cdot \tilde{I} + I_N \sigma^2 \quad (29)$$

where $\alpha_k$ is a constant greater than or equal to $-1$ and $e_k \in \mathbb{C}^N$ is a vector of all zeros but for $e_k(k) = 1$. For example, $\alpha_k = -1$ turns the entry $k$ of $\tilde{I}$ into 0, corresponding to a complete failure at node $k$.

Denote $y(t) = \Omega_{\tilde{V}}^{-1/2} \cdot \tilde{V}$ and $y'(t) = \Omega_{\tilde{V}}^{-1/2} \cdot \tilde{V}'$. Their corresponding covariance matrix can be calculated as:

$$\begin{aligned} E[(y(t) - \mu_{y(t)})(y(t) - \mu_{y(t)})^*] &= I_N \\ E[(y'(t) - \mu_{y'(t)})(y'(t) - \mu_{y'(t)})^*] &= I_N + [(1+\alpha_k)^2 - 1] \cdot \Omega_{\tilde{V}}^{-1/2} Z e_k \cdot e_k^* Z^* \Omega_{\tilde{V}}^{-1/2} \\ &\triangleq I_N + P(k, \alpha_k) \end{aligned} \quad (30)$$

where the $P(k, \alpha_k)$ in the second formular can be viewed as a perturbation into the identity matrix $I_N$. In particular, the perturbation part corresponds to the value change at the $k^{\text{th}}$ entry of the covariance matrix caused by the event occurrence at this node. By detecting and classifying this value change in the voltage covariance matrix, different types of events can be detected and classified.

III. NUMERICAL RESULTS

**Table 1. Detection criteria under different event occurrence scenarios in different system.**

| Test Feeder | Criterion | Scenario $\mathcal{H}_0$ [a] | FLT Events: | GL Events: | LI Events: | LS Events: | SC Events: | SA Events: | LT Events: | DG Events: |
|---|---|---|---|---|---|---|---|---|---|---|
| IEEE 34-node system | $C_{SRL}$ | 0.910 | 0.681 | 0.754 | 0.727 | 0.751 | 0.755 | 0.408 | 0.408 | 0.678 |
|  | $C_{MPL1}$ | 1.361 | 1.00E+06 | 1.782 | 2.397 | 2.841 | 1.822 | 1.15E+07 | 8.39E+06 | 3.601 |
|  | $C_{MPL2}$ | 0.140 | 0.515 | 0.306 | 0.302 | 0.312 | 0.204 | 0.594 | 0.460 | 0.306 |
| IEEE 37-node system | $C_{SRL}$ | 1.000 | 0.686 | 0.827 | 0.800 | 0.772 | 0.719 | 0.428 | 0.624 | 0.831 |
|  | $C_{MPL1}$ | 1.246 | 1.06E+06 | 1.948 | 2.346 | 1.781 | 1.703 | 6.20E+06 | 2.88E+05 | 1.968 |
|  | $C_{MPL2}$ | 0.120 | 0.305 | 0.289 | 0.261 | 0.305 | 0.317 | 0.399 | 0.364 | 0.303 |
| IEEE 123-node system | $C_{SRL}$ | 0.997 | 0.897 | 0.911 | 0.917 | 0.931 | 0.942 | 0.468 | 0.716 | 0.932 |
|  | $C_{MPL1}$ | 1.240 | 1.25E+05 | 1.519 | 1.552 | 1.589 | 1.893 | 4.53E+08 | 2.22E+06 | 1.477 |
|  | $C_{MPL2}$ | 0.102 | 0.266 | 0.168 | 0.159 | 0.169 | 0.171 | 0.658 | 0.498 | 0.192 |
| IEEE LV system | $C_{SRL}$ | 0.992 | 0.774 | 0.912 | 0.932 | 0.921 | 0.912 | 0.676 | 0.781 | 0.898 |
|  | $C_{MPL1}$ | 1.260 | 2.57E+04 | 3.505 | 3.733 | 1.784 | 1.756 | 3.81E+07 | 6.40E+07 | 1.864 |
|  | $C_{MPL2}$ | 0.109 | 0.366 | 0.169 | 0.172 | 0.174 | 0.171 | 0.577 | 0.445 | 0.168 |
| IEEE 8500-node system | $C_{SRL}$ | 1.000 | 0.960 | 0.986 | 0.986 | 0.987 | 0.992 | 0.952 | 0.972 | 0.982 |
|  | $C_{MPL1}$ | 1.012 | 1.25E+07 | 3.532 | 4.155 | 4.357 | 6.319 | 5.31E+08 | 2.21E+07 | 4.177 |
|  | $C_{MPL2}$ | 0.012 | 0.047 | 0.034 | 0.037 | 0.028 | 0.034 | 0.084 | 0.050 | 0.029 |
| Ckt-5 system | $C_{SRL}$ | 0.995 | 0.904 | 0.992 | 0.991 | 0.961 | 0.984 | 0.887 | 0.924 | 0.947 |
|  | $C_{MPL1}$ | 1.015 | 1.24E+05 | 3.736 | 3.989 | 3.629 | 6.423 | 9.98E+07 | 8.83E+07 | 3.936 |
|  | $C_{MPL2}$ | 0.013 | 0.049 | 0.017 | 0.018 | 0.020 | 0.020 | 0.069 | 0.056 | 0.019 |
| Ckt-24 | $C_{SRL}$ | 1.000 | 0.957 | 0.981 | 0.983 | 0.980 | 0.976 | 0.927 | 0.937 | 0.986 |

| | | | | | | | | | |
|---|---|---|---|---|---|---|---|---|---|
| system | $C_{MPL1}$ | 1.024 | 1.25E+07 | 3.519 | 3.552 | 5.357 | 4.319 | 4.53E+07 | 2.22E+06 | 4.774 |
| | $C_{MPL2}$ | 0.012 | 0.047 | 0.030 | 0.029 | 0.022 | 0.034 | 0.058 | 0.050 | 0.032 |

[a] Scenario $\mathcal{H}_0$: Scenario before event occurrence.